\documentclass{article}
\usepackage{graphicx}
\usepackage{amsmath}

\newtheorem{theorem}{Theorem}

\newtheorem{lemma}[theorem]{Lemma}

\begin{document}

\title{Schr\"{o}dinger operators with matrix potentials. Transition
from the
absolutely continuous to the singular spectrum.}
\author{S. Molchanov, B. Vainberg \and Dept. of Mathematics, University
of North
Carolina at Charlotte, \and Charlotte, NC 28223, USA}
\date{}
\maketitle

\begin{abstract}
It is proven that the absolutely continuous spectrum of matrix
Schr\"{o}\-ding\-er operators coincides (with the multiplicity
taken into account) with the spectrum of the unperturbed operator
if the (matrix) potential is square integrable. The same result is
also proven for some classes of slower decaying potentials if they
are smooth.

\end{abstract}

\textbf{1. Introduction.} The goal of this work is to prove the
''spectral $ L^{2}-$ conjecture'' for Schr\"{o}dinger operators
with matrix potentials. We shall consider the Hamiltonian
\begin{equation}
H\psi =-\frac{d^{2}\psi }{dx^{2}}+v(x)\psi ,\text{ \ \ \ }x\geq
0,\text{\ }
\label{nw1}
\end{equation}
acting in the space $L^{2}(R_{+})$ of vector functions $\psi (x)=[\psi
_{1},...,\psi _{n}]^{t}(x)$ with the boundary condition $\psi ^{\prime
}(0)=0.$ The potential $v$ here is a symmetric $n\times n$ matrix.
Since the
absolutely continuous (a.c.) spectrum of $H$ does not depend on the
type of
the boundary condition (b.c.), we restrict ourself to the case of the
Neumann b.c.. We could consider simultaneously operators on the half
axis
and on the whole axis. The multiplicity of the spectrum will be doubled
in
the second case. For the sake of simplicity we decided to focus on the
case
of the semiaxis. The results below can be also easily extended to the
case
of general canonical systems and lattice operators.

The spectral $L^{2}-$ conjecture concerns the minimal decay of the
potential
at infinity which still guarantees the existence of the rich a.c.
spectrum
of the operator. S. Kotani, N. Ushiroya \cite{ku} and F. Delyon, B.
Simon,
B. Souillard \cite{dss} described the bifurcation from a.c. to pure
point
(p.p.) spectrum for scalar Schr\"{o}dinger operators with modulated
(decaying) random potentials of the form:
\begin{equation*}
v(t,\omega )=\frac{\xi (t,\omega )}{1+\left| t\right| ^{\alpha }},
\end{equation*}
were $\xi (t,\omega ),$ $t\in R,$ is a Markov homogeneous ergodic
bounded
process on the probability space $(\Omega ,F,P)$. They proved that the
random Schr\"{o}dinger operator $H(\omega
)=-\frac{d^{2}}{dt^{2}}+v(t,\omega
)$ in $L^{2}$ on the whole axis has dense p.p. spectrum on $[0,\infty
)$
with probability one ($P-$ a.s.) if $\alpha <1/2,$ and its spectral
measure
on $[0,\infty )$ is $P-$ a.s. pure a.c. with multiplicity two if
$\alpha
>1/2.$ It turns out that the same type of bifurcation is valid in the
deterministic case and for potentials which are not necessarily
decaying as a power. This bifurcation is a consequence of the
spectral $L^{2}-$ conjecture which states that if $v\in L^{2}(R)$
then $\Sigma _{ac}(H)=[0,\infty )$ and the a.c. component $\mu
_{ac}$ of the spectral measure of operator $H$ is essentially
supported on $[0,\infty ),$ i.e. $\mu _{ac}(\Gamma )>0$ for any
Borel set $\Gamma ,$ $\left| \Gamma \right|
>0.$
Note that $v\in L^{2}(R)$ if $\alpha <1/2$ for the random potential
above,
and $v\notin L^{2}(R)$ if $\alpha >1/2.$ So, these results on random
operators show the exactness of the conjecture.

The spectral $L^{2}-$ conjecture for the scalar Schr\"{o}dinger
operators was justified in 1999 by Deift and Killip \cite{dk}. In
2001, the authors of this paper offered a different approach
\cite{mnv} which allowed us to get a sequence of conditions on the
potential $v$ such that under each of them the essential support
of the a.c. spectrum of the operator $H$ on the semiaxis $ R_{+}$
is $[0,\infty ).$ These conditions are related to the boundedness
of the first KdV integrals, and the first of those conditions is
$v\in L^{2}(R_{+}).$ Other conditions allow the potential to decay
slower if it is smooth.

This paper is devoted to the extension of our results for scalar
operators to the matrix Schr\"{o}dinger operators. We shall show
that the spectral measure of the operator (\ref{nw1}) has the a.c.
component, which has the multiplicity $n$ and is essentially
supported on $[0,\infty )$, if for some $ p\geq 0$ the following
functional is bounded:
\begin{equation}
J_{p}(v):=\int_{-\infty }^{\infty }\left(
||v^{(p-1)}(x)||^{2}+||v(x)||^{p+1}\right) dx,  \label{jp}
\end{equation}
It is assumed here that $v$ is extended by zero for $x<0$. This
implies that $v^{(j)}(0)=0$ for $j<p-1$. In fact, the latter
restriction is not essential, and one can easily show that the
main result remains valid if the lower limit in (\ref{jp}) is
replaced by zero. Examples of the condition (\ref{jp}) are
\begin{eqnarray*}
||v(x)|| &\in &L^{1},\text{ \ \ \ }p=0;\text{ \ \ \ }||v(x)||\in
L^{2},\text{ \ \ \ }p=1;
\end{eqnarray*}
\begin{eqnarray*}
\int_{-\infty }^{\infty }\left(
||\overset{.}{v}(x)||^{2}+||v(x)||^{3} \right) dx &<&\infty
,\text{ \ \ \ }p=2,\text{ \ \ etc}
\end{eqnarray*}

The result above is the simplest form of the generalized
$L^{2}-$conjecture for matrix operators. The case when
$v(x)=L+v_{0}(x),$ $L$ is a constant matrix, $\left\|
v_{0}\right\| \in L^{2}(R_{+}),$ will be published elsewhere. The
main feature of the latter model is the different multiplicity of
the a.c. component on the different intervals of the spectral
axis. These intervals are defined by the eigenvalues of the matrix
$ L.$

The main scheme of the proof of the matrix $L^{2}-$conjecture remains
the
same as in our work for the scalar case. We approximate the operator
$H$ by
operators $H_{s}$ with matrix potentials $v_{s}\in C_{0}^{\infty
}(R_{+})$
such that $v_{s}\rightarrow v$ in $L^{2}(R_{+})$ as $s\rightarrow
\infty ,$
and we construct the spectral measure $\mu (d\lambda )$ of operator $H$
as a
weak limit of spectral measures $\mu _{s}(d\lambda )$ of operators
$H_{s}.$
Then we express the measures $\mu _{s}(d\lambda )$ through the
scattering
data and use trace identities to prove that the limiting measure $\mu
(d\lambda )$ has the a.c. component of the same type as measures $\mu
_{s}(d\lambda ).$ However, the implementation of this scheme in the
matrix
case requires to overcome some difficulties related to
non-commutativity of
matrix multiplication. We shall also simplify some arguments used in
(\cite
{mnv}). So, we hope that this publication will make the proof much
more
transparent even in the scalar case.

In the next two sections of this paper we shall give\ a review of some
preliminary facts which we need for the proof, and we shall construct
the
spectral measure for the operator (\ref{nw1}) as a limit of spectral
measures of operators with compactly supported potentials. Until a
specific
condition on the potential is imposed, we assume that the potential
satisfies the Birman condition:
\begin{equation}
\int_{x}^{x+1}\left\| v(z)\right\| dz\leq c_{0}<\infty \ \text{\ \
for all } x\geq 0,  \label{bir}
\end{equation}
which implies that the spectrum of $H$ is bounded from below, and
therefore, the operator $H$ is essentially self adjoint. Most of
the results included in the sections 2 and 3 can be found in
\cite{a}, \cite{carl}, \cite{cl}, \cite{ks}, \cite{ls}.

The main results will be proved in the sections 4 and 5 of the paper.

\textbf{2. Symplectic structure. Green's matrix.} We assume here that
(\ref
{bir}) holds. The equation
\begin{equation}
H\psi =-\overset{..}{\psi }+v\psi =\lambda \psi ,\qquad x\in R_{+},
\label{1}
\end{equation}
can be written in the canonical form. Put $\psi ^{\prime }=p$ and
$Y=(\psi
,p)^{t},$ then
\begin{equation}
-J\overset{.}{Y}=VY+\lambda QY,  \label{2}
\end{equation}
where
\begin{equation}
J=\left[
\begin{array}{cc}
0 & -I \\
I & 0
\end{array}
\right] ,\text{ \ \ \ }V=\left[
\begin{array}{cc}
v & 0 \\
0 & -I
\end{array}
\right] ,\qquad Q=\left[
\begin{array}{cc}
I & 0 \\
0 & 0
\end{array}
\right] .  \label{s3}
\end{equation}
Here $I$ is the identity $n\times n$ matrix, $0$ is $n\times n$ matrix
of
zeroes and $-J^{2}$ is the identity $2n\times 2n$ matrix.

We denote by $M_{\lambda }(0,x)$ the transfer matrix for the system
(\ref{2}
) (or (\ref{1})) which is given by matrix equation
\begin{equation}
-J\overset{.}{M_{\lambda }}=(V+\lambda Q)M_{\lambda },\text{ \ \
}M_{\lambda
}(0,0)=I.  \label{s4}
\end{equation}
It is well known that $M_{\lambda }(0,x)\in Sp(2n)\subset SL(2n,R).$
The
symplectic group $Sp(2n)$ consists of matrices $M$ which preserve the
skew-symmetric product $\left\langle x,y\right\rangle =(Jx,y),$ i.e.
$M\in
Sp(2n)$ if $\left\langle Mx,My\right\rangle =\left\langle
x,y\right\rangle .$
An equivalent characteristics of symplectic matrices is the identity
\begin{equation}
M^{t}JM=J.  \label{s5}
\end{equation}
One can easily check that $M_{\lambda }(0,x)$ satisfies (\ref{s5})
by differentiating both sides of (\ref{s5}) with $M=M_{\lambda }$
and using (\ref{s4}) and (\ref{s3}). Thus, for any $x,$
\begin{equation}
M_{\lambda }(0,x)\in Sp(2n).  \label{sp1}
\end{equation}

If $M\in Sp(2n)$ is presented in the block form
\begin{equation*}
M=\left[
\begin{array}{cc}
P & Q \\
R & T
\end{array}
\right]
\end{equation*}
with $n\times n$ matrices $P,Q,R,T,$ then the identity (\ref{s5})
implies
''pseudo commutativity'':
\begin{equation*}
P^{t}R=R^{t}P,\text{ \ \ }T^{t}Q=Q^{t}T
\end{equation*}
and the ''matrix unimodularity'':
\begin{equation*}
T^{t}P-Q^{t}R=I.
\end{equation*}
Since $M^{t}\in Sp(2n)$ then also
\begin{equation*}
PQ^{t}=QP^{t},\text{ \ \ }RT^{t}=TR^{t}.
\end{equation*}

We shall need the concept of the Lagrangian plane. Linear
$n$-dimensional subspace $\pi \in R^{2n}$ is called the Lagrangian
plane if $\left\langle x,y\right\rangle =(Jx,y)=0$ for any $x,y\in
\pi .$ If $\pi $ is a Lagrangian plane and $M\in Sp(2n)$ then
$M\pi $ is also a Lagrangian plane. In particular, the transfer
matrix maps any Lagrangian plane into a Lagrangian plane. It will
be used in the following context. Let $u_{\lambda }(x)$ be a $
n\times n$ matrix solution of the system (\ref{1}). Consider its
Cauchy data $[u_{\lambda },\overset{.}{u}_{\lambda }]^{t}.$ Let
$\pi (x)$ be the span of the columns of the latter matrix. If $\pi
(x)$ is a Lagrangian plane for one value of $x$ then $\pi (x)$ is
a Lagrangian plane for any $x$.

The Wronskian of two $n\times n$ matrices is a $n\times n$ matrix
defined by
the formula
\begin{equation}
W(u,v)=\left|
\begin{array}{cc}
u(x) & v(x) \\
\overset{.}{u}(x) & \overset{.}{v}(x)
\end{array}
\right| =\overset{.}{v}^{t}(x)u(x)-v^{t}(x)\overset{.}{u}(x).
\label{wr}
\end{equation}
The Wronskian can also be written in the form
\begin{equation}
W(u,v)=[v^{t},\overset{.}{v}^{t}]J[
\begin{array}{c}
u \\
\overset{.}{u}
\end{array}
].  \label{nwr}
\end{equation}
The same arguments as in the scalar case imply that the Wronskian
$W(u,v)$
does not depend on $x$ if the matrices $u,v$ satisfy the equation
(\ref{1}).

Let us consider the following Sturm-Liouville problem for the equation
(\ref
{1})
\begin{equation}
(H-\lambda )\psi =0,\text{ \ \ }x_{1}<x<x_{2};\text{ \ }[\psi
,\overset{.}{ \psi }]_{x=x_{1}}^{t}\in \pi _{1},\text{ \ }[\psi
,\overset{.}{\psi } ]_{x=x_{2}}^{t}\in \pi _{2},  \label{sl}
\end{equation}
where $\pi _{1},$ $\pi _{2}$\ are two Lagrangian planes.
Conservation of the Wronskian allows us to construct the Green
function for that problem. Let $ u_{\lambda }$ $(v_{\lambda })$ be
a matrix whose columns are solutions of the Cauchy problem for the
equation (\ref{1}) with the Cauchy data at $x=x_{1}$ ($x=x_{2})$
which form a basis in $\pi _{1}$ ($\pi _{2},$ respectively).

\begin{lemma}
\label{nl2}The determinant $\det W(u_{\lambda },v_{\lambda })$ is equal
to
zero if and only if $\lambda $ is an eigenvalue of the problem
(\ref{sl}).
If $\det W\neq 0$ then the matrix
\begin{equation*}
G_{\lambda }(x,\xi )=\left\{
\begin{array}{c}
-u_{\lambda }(x)[W(u_{\lambda },v_{\lambda })]^{-1}v_{\lambda
}^{t}(\xi ),
\text{ \ \ }x\leq \xi , \\
-v_{\lambda }(x)[W^{t}(u_{\lambda },v_{\lambda })]^{-1}u_{\lambda
}^{t}(\xi
),\text{ \ \ }x>\xi ,
\end{array}
\right.
\end{equation*}
is the Green matrix for the problem (\ref{sl}).
\end{lemma}

Let us consider the $2n\times 2n$ matrix $\widetilde{W}$ which can
be found in the middle of equalities (\ref{wr}). The space $\pi
_{1}(x)$ ($\pi _{2}(x))$ spanned by the first (respectively, last)
$n$ columns of $ \widetilde{W}$ is a Lagrangian plane, and
therefore $\det W(u_{\lambda },v_{\lambda })=\det \widetilde{W}$
(it also follows from (\ref{st})). Thus, $\det W(u_{\lambda
},v_{\lambda })=0$ if and only if $\pi _{1}(x)$ and $\pi _{2}(x)$
have a nontrivial intersection, in particular, for $x=x_{2}.$ The
latter is equivalent to the existence of a nontrivial solution of
(\ref{sl} ). The first statement of the Lemma is proved.

In order to prove the second statement of the Lemma one needs only to
show
that the matrix $G_{\lambda }(x,\xi )$ is continuous at $x=\xi $ and
its
derivative has a jump equal to $-I.$ If $\pi $ is a Lagrangian plane
then
planes $\pi $ and $J\pi $ are orthogonal. From here and (\ref{nwr}) it
follows that $W(u_{\lambda },u_{\lambda })$ and $W(v_{\lambda
},v_{\lambda
}) $ are zero matrices, and
\begin{equation}
\left( \widetilde{W}\right) ^{t}J\widetilde{W}=\left[
\begin{array}{cc}
0 & -W^{t}(u_{\lambda },v_{\lambda }) \\
W(u_{\lambda },v_{\lambda }) & 0
\end{array}
\right] .  \label{st}
\end{equation}
This implies that
\begin{equation*}
\widetilde{W}\left[
\begin{array}{cc}
0 & -[W(u_{\lambda },v_{\lambda })]^{-1} \\
-[W^{t}(u_{\lambda },v_{\lambda })]^{-1} & 0
\end{array}
\right] \left( \widetilde{W}\right) ^{t}=-J.
\end{equation*}
The first column of this matrix relation gives the necessary
properties of $ G_{\lambda }$ at $x=\xi .$ \ The Lemma is proven.

\textbf{3. Spectral measure. }We still assume that the Birman
condition (\ref {bir}) holds. As it was mentioned earlier, this
implies the boundedness of the spectrum of $H$ from below:

\begin{lemma}
\label{nls} If the Birman condition (\ref{bir}) holds then there
exists $\Lambda _{0}>-\infty $ such that $\Sigma (H)\subset
\lbrack \Lambda _{0},\infty ).$
\end{lemma}

\textbf{Proof. }Obviously, it is sufficient to prove that
\begin{equation*}
(H\varphi ,\varphi )\geq \Lambda _{0}>-\infty
\end{equation*}
for all $\varphi $ such that
\begin{equation*}
\varphi \in C_{0}^{\infty }(R_{+}),\text{ \ }\overset{.}{\varphi
}(0)=0, \text{ \ }\left\| \varphi \right\| _{L_{2}}=1.
\end{equation*}
From the standard Neumann-Dirichlet estimation it follows that
$\Lambda _{0}$ can only increase if we allow $\varphi $ to have
jumps at integer points $ x=n\geq 0,$ but impose the Neumann b.c.
at those points. To be more exact, it is enough to show that there
exists $\Lambda _{0}>-\infty $ such that, for any $n$ and any
smooth function $\varphi $ on the interval $\Delta _{n}=(n,n+1),$
\begin{equation}
(H_{n}\varphi ,\varphi ):=\int_{n}^{n+1}[(\overset{.}{\varphi
},\overset{.}{ \varphi })+(v\varphi ,\varphi )]dx\geq \Lambda _{0}
\label{22}
\end{equation}
if
\begin{equation*}
\overset{.}{\varphi }(n)=\overset{.}{\varphi }(n+1)=0,\text{ \
}\left\|
\varphi \right\| _{L_{2}(\Delta _{n})}=1.
\end{equation*}

Since $\left\| \varphi \right\| _{L_{2}(\Delta _{n})}=1,$ there exists
a
point $x_{0}\in \Delta _{n}$ such that $\left| \varphi (x_{0})\right|
=1.$
Then
\begin{equation*}
\left| \varphi (x)\right| ^{2}-\left| \varphi (x_{0})\right|
^{2}=2\int_{x_{0}}^{x}(\varphi ,\overset{.}{\varphi })dz,
\end{equation*}
and therefore, for any $\varepsilon >0,$
\begin{equation*}
\left| \varphi (x)\right| ^{2}\leq 1+2\int_{n}^{n+1}|(\varphi
,\overset{.}{ \varphi })|dz\leq 1+\varepsilon \left\| \varphi
\right\| _{L_{2}(\Delta _{n})}^{2}+\varepsilon ^{-1}\left\|
\overset{.}{\varphi }\right\| _{L_{2}(\Delta _{n})}^{2}.
\end{equation*}
Now,
\begin{eqnarray*}
\left| \int_{n}^{n+1}(v\varphi ,\varphi )dx\right| &\leq
&\int_{n}^{n+1}\left\| v(z)\right\| \left| \varphi (z)\right| ^{2}dz
\\
&\leq &(1+\varepsilon +\varepsilon ^{-1}\left\|
\overset{.}{\varphi }\right\| _{L_{2}(\Delta
_{n})}^{2})\int_{n}^{n+1}\left\| v(z)\right\| dz.
\end{eqnarray*}
Finally,
\begin{equation*}
(H_{n}\varphi ,\varphi )\geq \left\| \overset{.}{\varphi }\right\|
_{L_{2}(\Delta _{n})}^{2}-(1+\varepsilon +\varepsilon ^{-1}\left\|
\overset{. }{\varphi }\right\| _{L_{2}(\Delta
_{n})}^{2})\int_{n}^{n+1}\left\| v(z)\right\| dz.
\end{equation*}
This inequality with
\begin{equation*}
\varepsilon =\int_{n}^{n+1}\left\| v(z)\right\| dz
\end{equation*}
implies (\ref{22}) with
\begin{equation*}
\Lambda _{0}=-(1+\int_{n}^{n+1}\left\| v(z)\right\| dz).
\end{equation*}

The proof is completed.

After the boundedness of the spectrum of $H$ is established, the
general
theory provides the essential self adjointness of $H,$ analyticity of
the
resolvent $R_{z}=(H-z)^{-1},$ $z\notin \lbrack \Lambda _{0},\infty ),$
the
existence, for any element $\varphi \in L^{2}(R_{+}),$ of the spectral
measures $\mu _{\varphi }$ such that
\begin{equation*}
(R_{z}\varphi ,\varphi )=\int_{\Lambda _{0}}^{\infty }\frac{\mu
_{\varphi
}(d\lambda )}{\lambda -z},
\end{equation*}
etc. This paper concerns a very special spectral measure $\mu (d\lambda
)$
associated to the generalized Fourier transform defined by operator
$H.$
This measure contains information about all spectral measures $\mu
_{\varphi
}(d\lambda )$ and describes the spectral type of $H.$

The generalized Fourier transform $F$ for vector functions $\varphi
\in
L_{2}(R_{+})$ is defined by the formula
\begin{equation}
\widehat{\varphi }(\lambda )=F(\varphi (x))=\int_{0}^{\infty
}u_{\lambda
}^{t}(x)\varphi (x)dx.  \label{gf}
\end{equation}
where $u_{\lambda }$ is the matrix solution of the problem
\begin{equation*}
(H-\lambda )u_{\lambda }=0,\text{ \ }x>0;\text{ \ }u_{\lambda
}(0)=I,\text{
\ }\overset{.}{u}_{\lambda }(0)=0.
\end{equation*}
There exists a unique (due to self adjointness of $H$) spectral measure
$\mu
(d\lambda )$ such that the inverse Fourier transform is given by the
formula
\begin{equation*}
\varphi (x)=\int_{\Lambda _{0}}^{\infty }u_{\lambda
}(x)\widehat{\varphi } (\lambda )\mu (d\lambda )
\end{equation*}
and Parseval's identity holds
\begin{equation*}
(\varphi _{1}(x),\text{ }\varphi _{2}(x))=\int_{\Lambda
_{0}}^{\infty }( \widehat{\varphi _{1}}(\lambda ),\widehat{\varphi
_{2}}(\lambda ))\mu (d\lambda ).
\end{equation*}
One has to treat integrals here as $L_{2}$-limits of integrals
over the finite intervals $(\Lambda _{0},L)$ as $L\rightarrow
\infty .$ The mapping $ \varphi \rightarrow \widehat{\varphi }$ is
the isomorphism between $ L_{2}(R_{+},$ $dx)$ and $L_{2}(R,$ $\mu
(d\lambda )).$ The spectral measures $\mu _{\varphi }(d\lambda )$
of elements $\varphi \in L_{2}(R_{+})$ can be expressed through
$\mu (d\lambda )$ by the formula
\begin{equation}
\mu _{\varphi }(d\lambda )=\left| \widehat{\varphi }(\lambda )\right|
^{2}\mu (d\lambda ).  \label{nsm}
\end{equation}

One of the methods (attributed to B. Levitan) to prove the
statements above on the generalized Fourier transform and the
spectral measure is based on the weak convergence $\mu
_{k}\rightarrow \mu $ of the discrete measures $ \mu _{k}$
associated to the restriction of the Hamiltonian $H$ to the set of
functions on the interval $(0,k)\subset R_{+}$ with a Lagrangian
b.c. at the point $x=k.$ This method is very convenient if one
wants to prove that the spectral measure $\mu $ is discrete. In
order to describe the a.c. component of $\mu $ it is better to
approximate $\mu $ by a.c. measures $\mu _{k}.$ One of such
approximations (going back to M. Krein and his school) is based on
the averaging of the measures $\mu _{k}$ with respect to the
Lagrangian planes related to the b.c. at $x=k.$ We use a different
approximation of the spectral measure $\mu .$

Let $v_{s}$ be a sequence of symmetric matrix potentials such that
$v_{s}$
is supported on $[0,s]$ and
\begin{equation*}
\int_{0}^{s}||v_{s}(x)-v(x)||dx\rightarrow 0\text{ \ \ as \ \
}s\rightarrow
\infty .
\end{equation*}
We denote by $H_{s}$ the Hamiltonian on $L_{2}(R_{+})$ with the
potential $ v_{s}$ and the Neumann b.c. at $x=0.$ Let $\mu
_{s}(d\lambda )$ be the spectral measure of the operator $H_{s}.$
This measure is a.c. on $[0,\infty )$ (with multiplicity $n$) and
has at most a finite number of atoms on the negative semiaxis at
eigenvalues $\lambda _{j,s}<0,$ $j\leq m(s),$ of the operator
$H_{s}.$

\begin{lemma}
\label{nlsm} Spectral measures $\mu _{s}(d\lambda )$ converge weakly to
$\mu
(d\lambda )$ as $s\rightarrow \infty .$
\end{lemma}

The proof of the Lemma is based on the convergence of the
resolvent $ R_{z}^{(s)}=(H_{s}-z)^{-1}$ and of the corresponding
Weil's function when $ \mathrm{Im}z>0,$ $s\rightarrow \infty .$
Similar arguments were used in \cite {ks} in a different setting.
We shall provide the technical details of the proof of this lemma
elsewhere.

\textbf{4. Operators with smooth compactly supported potentials.} We
assume
in this section that $v\in C_{0}^{\infty }(R_{+}).$\ Let $u_{\lambda
}=u_{\lambda }(x)$ be the matrix solution of the problem
\begin{equation*}
\mathbf{(}H-\lambda )u=0,\text{ \ }x>0;\text{ \ }u(0)=I,\text{ \
}\overset{. }{u}(0)=0,
\end{equation*}
and let $v_{\lambda }^{+},$ $v_{\lambda }^{-}$ be matrix solutions
of the some equation such that $v_{\lambda }^{\pm }(x)=e^{\pm
i\sqrt{\lambda }x}I$ as $x>>1.$ If $\mathrm{Im}\lambda >0$ then
elements of the matrix $ v_{\lambda }^{+}$ ($v_{\lambda }^{-}$)
decay exponentially as $x\rightarrow \infty $ ($x\rightarrow
-\infty $) and they grow exponentially as $ x\rightarrow -\infty $
($x\rightarrow \infty $)$.$ The columns of the matrices
$v_{\lambda }^{+}$ $\ $and $v_{\lambda }^{-}$ taken together form
a basis in the solution space of the equation
$\mathbf{(}H_{s}-\lambda )u=0.$ Thus,
\begin{equation}
u_{\lambda }(x)=v_{\lambda }^{+}(x)A(\lambda )+v_{\lambda
}^{-}(x)B(\lambda
),  \label{nuv}
\end{equation}
where matrices $A(\lambda ),$ $B(\lambda )$ are analytic in $\lambda $
(with
a branch point at $\lambda =0),$ and
\begin{equation}
A=\overline{B}\text{ \ for }\lambda >0,  \label{l0}
\end{equation}
since $u_{\lambda }$ is real when $\lambda >0.$ Operator $H$ has at
most a
finite number of eigenvalues $\lambda _{j},$ $j\leq m,$ and $\lambda
_{j}<0.$

\begin{lemma}
\label{n1} The spectral measure $\mu (d\lambda )$ of the operator $H$
is
equal to
\begin{equation*}
\mu (d\lambda )=\overset{m}{\underset{j=1}{\sum }}\delta (\lambda
-\lambda
_{j})d\lambda +\frac{\theta (\lambda )}{4\pi \sqrt{\lambda }}N(\lambda
)d\lambda ,\text{ \ }N(\lambda )=[B(\lambda
)]^{-1}[\overline{B^{t}}(\lambda
)]^{-1}\text{\ },
\end{equation*}
where $\theta (\lambda )=1$ for $\lambda \geq 0,$ $\theta (\lambda
)=0$ for $ \lambda <0.$
\end{lemma}

\textbf{Proof.} We shall use below the following obvious properties of
the
Wronskians. If $u=u(x),$ $v=v(x)$ and $Q$ are arbitrary $n\times n$
matrices, and $Q$ does not depend on $x$, then
\begin{equation}
W(u,u)=0,\text{ \ }W(u,vQ)=Q^{t}W(u,v),\text{ \ \ }W(uQ,v)=W(u,v)Q.
\label{nfw}
\end{equation}
Note that
\begin{equation}
W(v^{+},v^{+})=0,\text{ \ \
}W(v^{-},v^{+})=-W(v^{+},v^{-})=2i\sqrt{\lambda } I.  \label{nvr}
\end{equation}
In order to justify these relations one can evaluate the
Wronskians for $x>>1 $ and use their independence of $x.$
Relations (\ref{nuv}), (\ref{nfw}) and (\ref{nvr}) immediately
imply that
\begin{equation}
W(u_{\lambda },v^{+})=-2i\sqrt{\lambda }A,\text{ \ \ }W(u_{\lambda
},v_{\lambda }^{-})=2i\sqrt{\lambda }B,  \label{n11}
\end{equation}
and
\begin{equation*}
0=W(u_{\lambda },u_{\lambda
})=W(v^{+}A,v^{-}B)+W(v^{-}B,v^{+}A)=2i\sqrt{ \lambda
}[A^{t}B-B^{t}A].
\end{equation*}
Thus,
\begin{equation}
A^{t}B=B^{t}A.  \label{naa}
\end{equation}

If $G=G_{\lambda }(x,\xi )$ is Green's matrix of the problem
\begin{equation*}
\mathbf{(}H-\lambda )G=\delta (x-\xi )I,\text{ \ }x>0,\text{ \
}\mathrm{Im} \lambda \neq 0;\text{ \ \ }\overset{.}{G}(0,\xi
)=G(\infty ,\xi )=0,
\end{equation*}
then from (\ref{n11}) and Lemma \ref{nl2}\ it follows that, for $x<\xi
,$

\begin{eqnarray*}
G_{\lambda }(x,\xi ) &=&\frac{-1}{2i\sqrt{\lambda }}u_{\lambda
}(x)B^{-1}[v^{+}(\xi )]^{t},\text{\ \ \ }\mathrm{Im}\lambda >0, \\
G_{\lambda }(x,\xi ) &=&\frac{1}{2i\sqrt{\lambda }}u_{\lambda
}(x)A^{-1}[v^{-}(\xi )]^{t},\text{\ \ \ }\mathrm{Im}\lambda <0.
\end{eqnarray*}
One could also write the corresponding formulas when $x>\xi .$
Thus, if $\lambda >0$ and $x<\xi ,$ then (\ref{naa}) and
(\ref{nuv}) imply
\begin{equation*}
G_{\lambda +i0}(x,\xi )-G_{\lambda -i0}(x,\xi )
\end{equation*}
\begin{equation*}
=\frac{-1}{2i\sqrt{\lambda }}u_{\lambda
}(x)B^{-1}(A^{-1})^{t}A^{t}[v^{+}(\xi
)]^{t}-\frac{1}{2i\sqrt{\lambda }} u_{\lambda
}(x)A^{-1}(B^{-1})^{t}B^{t}[v^{-}(\xi )]^{t}
\end{equation*}
\begin{equation*}
=\frac{-1}{2i\sqrt{\lambda }}u_{\lambda
}(x)B^{-1}(A^{-1})^{t}u_{\lambda
}(\xi ).
\end{equation*}
The same formula is valid when $x>\xi .$ One can derive it
similarly, but the easier way to get it is to note that the
operator with the kernel $i[G_{\lambda +i0}-G_{\lambda -i0}]$ is
symmetric. This and (\ref{l0}) imply
\begin{equation*}
G_{\lambda +i0}(x,\xi )-G_{\lambda -i0}(x,\xi
)=\frac{1}{2i\sqrt{\lambda }} u_{\lambda
}(x)B^{-1}(\overline{B}^{-1})^{t}u_{\lambda }(\xi ),\text{ \ \ \ }
\lambda >0.
\end{equation*}
To complete the proof of Lemma \ref{n1}\ it remains only to apply the
Stone
formula and use (\ref{nsm}).

Let us denote by $w=w_{\lambda }(x)$ the matrix \textit{Jost solution}
for
operator $H$. This solution is defined by the relations
\begin{equation*}
(H-\lambda )w=0,\text{ \ \ }x>0;\text{ \ \ }w(0)=I,\text{ \
}\overset{.}{w} (0)=-i\sqrt{\lambda }I\text{\ }.
\end{equation*}
Similarly to (\ref{nuv}) we have
\begin{equation}
w_{\lambda }(x)=v_{\lambda }^{+}(x)P(\lambda )+v_{\lambda
}^{-}(x)S(\lambda
),  \label{24a}
\end{equation}
where matrices $P(\lambda )$ and $S(\lambda )$ are analytic in
$\lambda $ with a branch point at $\lambda =0.$ If the Jost
solution is extended by $ e^{-i\sqrt{\lambda }x}I$ for $x<0$ and
$H^{\prime }$ is the Hamiltonian $H$ on the whole axis with the
potential $v$ extended by zero for $x<0,$ then $ [S(\lambda
)]^{-1}w_{\lambda }(x)$ is the \textit{scattering solution} for
the operator $H^{\prime }.$ It describes the propagation of the
incident plane wave $e^{-i\sqrt{\lambda }x}I$ coming from
$x=\infty $ with $ [S(\lambda )]^{-1}$ being the transmission
matrix and $P(\lambda )[S(\lambda )]^{-1}$ being the reflection
matrix. We shall call $S(\lambda )$ the \textit{Jost transmission
matrix}. Note that (\ref{24a}) for large enough $x$ can be written
in the form
\begin{equation}
w_{\lambda }(x)=e^{i\sqrt{\lambda }x}P(\lambda
)+e^{-i\sqrt{\lambda } x}S(\lambda ),\text{ \ \ }x>>1.
\label{26aa}
\end{equation}

The following theorem allows us to estimate the density of the
spectral measure $\mu (d\lambda )$ through the Jost transmission
matrix which has better asymptotic behavior for complex $\lambda
\rightarrow \infty $ than $B(\lambda ).$

\begin{theorem}
\label{tes}The following estimates hold for the matrix $N(\lambda
)$
\begin{equation*}
\frac{1}{4}\leq ||[N(\lambda )]^{-1}||\leq ||S(\lambda )||^{2}\leq
|\det
S(\lambda )|^{2}.
\end{equation*}
\end{theorem}

\textbf{Proof.} Green's formula for the columns of the matrix
$w=w_{\lambda
}(x)$ can be written in the form
\begin{equation*}
0=\int_{0}^{a}[(Hw)^{t}\overset{..}{w}-\overset{..}{w}^{t}Hw]dx=2i\mathrm{Im}
[(\overset{.}{w}^{t}\overline{w})(a)-(\overset{.}{w}^{t}\overline{w})(0)].
\end{equation*}
Thus,
\begin{equation}
\mathrm{Im}(\overset{.}{w}^{t}\overline{w})(x)=\sqrt{\lambda
}I,\text{ \ \ } x>0.  \label{26a}
\end{equation}

We choose $s$ so big that (\ref{26aa}) holds for $x>s.$ If we
substitute ( \ref{26aa}) into (\ref{26a}), take the average of
both sides of (\ref{26a}) over interval $(s,s+l)$ and pass to the
limit as $l\rightarrow \infty ,$ then we arrive to the following
relation (''conservation of the energy''):
\begin{equation}
\overline{S}^{t}(\lambda )S(\lambda )-\overline{P}^{t}(\lambda
)P(\lambda
)=I,\text{ \ \ \ }\lambda >0\text{.}  \label{ab}
\end{equation}
In particular, from (\ref{ab}) it follows that $||S\varphi ||\geq
||\varphi ||$ for any vector $\varphi ,$ and therefore,
\begin{equation}
|\mu _{j}(\lambda )|\geq 1,\text{ \ \ \ }\lambda >0\text{,}
\label{abc}
\end{equation}
where $\mu _{j}(\lambda )$ are eigenvalues of the matrix $S(\lambda
).$
Formulas (\ref{ab}) and (\ref{abc}) imply that
\begin{equation}
1+||P||\leq ||S||\text{ \ \ and }||S||\leq |\det S|,\text{ \ \ \
}\lambda >0 \text{.}  \label{27a}
\end{equation}

Obviously, $u_{\lambda }=\frac{1}{2}(w_{\lambda }+\overline{w}_{\lambda
}),$
$\lambda >0.$ Thus, from (\ref{nuv}) and (\ref{24a}) it follows that
\begin{equation*}
B=\frac{1}{2}(S+\overline{P}),\text{ \ \ \ }\lambda >0.
\end{equation*}
From here and (\ref{27a}) it follows that
\begin{equation*}
||\overline{B}^{t}B||=\frac{1}{4}||S+\overline{P}||^{2}\leq
||S||^{2}\leq
|\det S|,
\end{equation*}
and
\begin{equation*}
||\overline{B}^{t}B||=\frac{1}{4}||S+\overline{P}||^{2}\geq
\frac{1}{4} (||S||^{2}-||\overline{P}||)^{2}\geq \frac{1}{4}.
\end{equation*}
The proof is completed.

We shall need an asymptotic expansion of det$S(\lambda )$ at
infinity. Let us recall that $S(\lambda )$ is analytic in the
complex $\lambda $-plane $ \mathbf{C}$ with a branch point at the
origin. Let $\mathbf{C}_{1}=\mathbf{C\backslash }[0,\infty ).$

Let $P=P(v,\overset{.}{v},...)$ be a polynomial of $v$ and its
derivatives. Note that terms of $P$ depend on the order of their
factors. We shall say that $P$ is generalized homogeneous of order
$m$ if the substitution $v^{(l)}(x)\rightarrow \varepsilon
^{2+l}v^{(l)}(x)$ in the arguments of $P$ results in
multiplication of $P$ by $\varepsilon ^{m}.$

\begin{theorem}
\label{t8}If the matrix potential $v$ belongs to $ C_{0}^{\infty
}(R_{+})$ then the following expansion is valid
\begin{equation}
\ln [\det S(\lambda )]\sim i\overset{\infty }{\underset{m=0}{\sum
}}\frac{ I_{m}}{\lambda ^{m+1/2}},\text{ \ \ }\lambda \in
\mathbf{C}_{1},\text{ \ \ } |\lambda |\rightarrow \infty ,
\label{ds}
\end{equation}
where $I_{m}=I_{m}(v)$ are functionals of $v$ of the form
\begin{equation}
I_{m}=\int_{0}^{\infty }P_{m}(v,\overset{.}{v},...)dx.  \label{pm}
\end{equation}
Here $P_{m}$ are generalized homogeneous polynomials of $v$ and
its derivatives, and the order of $P_{m}$ is $2m+2.$
\end{theorem}

\textbf{Remarks}. 1). Polynomials $P_{m}$ are not defined
uniquely. For example, one can add $\overset{.}{v}$ to any
$P_{m}.$ After polynomials $P_{m}$ are found, one can change them
in such a way (by integrating by parts in (\ref{pm})) that $P_{m}$
will depend on derivatives of $v$ of the order at most $m-1.$

2). The first polynomials $P_{m}$ are:
\begin{equation*}
P_{0}=v,\text{ \ }P_{1}=v^{2},\text{ \
}P_{2}=\frac{1}{2}\overset{.}{v} ^{2}+v^{3},\text{ \ \
}P_{3}=\frac{1}{2}\overset{..}{v}^{2}-\frac{5}{4}v^{2}
\overset{..}{v}-\frac{5}{4}\overset{..}{v}v^{2}+\frac{5}{4}v^{4}.
\end{equation*}

3). Theorem \ref{t8} is well known in the scalar case. In this case,
$I_{m}$
are first integrals of the KdV equation.

\textbf{Proof.} Put $w=e^{-i\sqrt{\lambda }x}z.$ Then
\begin{equation}
-\overset{..}{z}+2i\sqrt{\lambda }\overset{.}{z}+v(x)z=0,\text{ \
\ }x\in R; \text{ \ \ }z=I\text{ \ for \ }x<0.  \label{ss}
\end{equation}

We denote $\overset{.}{z}z^{-1}$ by $Q(x,\lambda ).$ Then
$\overset{.}{z}=Qz$ and
$\overset{..}{z}=\overset{.}{Q}z+Q\overset{.}{z}=(\overset{.}{Q}
+Q^{2})z. $ From here and (\ref{ss}) it follows that
\begin{equation*}
-\overset{.}{Q}-Q^{2}+2i\sqrt{\lambda }Q+v(x)=0,\text{ \ \ }x\in
R;\text{ \ } Q=0\text{ \ for \ }x<0.
\end{equation*}

Let $v=0$ for $x>s.$ If $|x|\leq s+1,$ $\lambda \in
\mathbf{C}_{1},$ $|\lambda |\rightarrow \infty ,$ then solution
$Q$ of the above matrix equation can be easily found in the form
of the power series:
\begin{equation}
Q\sim \overset{\infty }{\underset{1}{\sum }}(2i\sqrt{\lambda
})^{-m}Q_{m}(x), \text{ \ \ }|x|\leq s+1,\text{ }\lambda \in
\mathbf{C}_{1},\text{ }|\lambda |\rightarrow \infty ,  \label{5s}
\end{equation}
where
\begin{equation*}
Q_{1}=v,\text{ \ \ }Q_{2}=\overset{.}{v},\text{ \
}Q_{m}(x)=\overset{.}{Q} _{m-1}+\overset{m-2}{\underset{k=1}{\sum
}}\text{\ }Q_{k}\text{\ }Q_{m-k-1}, \text{ \ \ }m>2.
\end{equation*}
One can show by induction that $Q_{m}$ are generalized homogeneous
polynomials of $v$ and its derivatives, and ord$Q_{m}=m+1.$

In order to find $z$ for $|x|\leq s+1$ one has
 to solve the following
matrix equation
\begin{equation}
\overset{.}{z}=Qz,\text{ \ \ }x\in R;\text{ \ \ }z=I\text{ \ for \
}x<0.
\label{1s}
\end{equation}
We shall
 forget temporary about dependence of $Q$ on $\lambda .$
The solution of (\ref{1s}) can be written in the form of the
matrix multiplicative integral:
\begin{equation}
z(x)=\overset{x}{\underset{t=1}{\Pi }}(I+Q(t)dt),  \label{2s}
\end{equation}
which is a short notation for the following limit
\begin{equation}
z(x)=\underset{\max |\Delta x_{i}|\rightarrow 0}{\lim }\text{
}\overset{m}{ \underset{i=1}{\Pi }}(I+Q(x_{i})\Delta x_{i}).
\label{3s}
\end{equation}
Here $x_{i},$ $0\leq i\leq m,$ are points on $[0,x],$
$x_{i}<x_{i+1},$ $x_{0}=0,$ $x_{m}=x,$ and $\Delta
x_{i}=x_{i+1}-x_{i}.$ Thus, one arrives to the multiplicative
integral (\ref{2s}) if (\ref{1s}) is solved by the Euler method.
From (\ref{2s}), (\ref{3s}) it follows that
\begin{equation*}
\det z(x)=\overset{x}{\underset{t=1}{\Pi }}(I+\text{tr}Q(t)dt).
\end{equation*}
The last expression is the solution of the scalar equation
\begin{equation*}
\overset{.}{y}=[\text{tr}Q(x)]y,\text{ \ \ }x\in R;\text{ \ \
}y(x)=1\text{
\ for \ }x<0.
\end{equation*}
Thus,
\begin{equation}
\det z(x)=e^{\int_{0}^{\infty }\text{tr}Q(t)dt}\text{.}  \label{4s}
\end{equation}

Since $Q_{m}$ are polynomials of $v$ and its derivatives, then
$Q_{m}=0$ for $x>s,$ that is
 $Q=O(|\lambda |^{-\infty })$ when $s\leq x\leq s+1,$
$\lambda
\in \mathbf{C}_{1},$ $|\lambda |\rightarrow \infty .$ Then (\ref{1s})
implies that
\begin{equation*}
z(x,\lambda )=z_{0}(\lambda )+O(|\lambda |^{-\infty }),\text{ \ \
}s\leq
x\leq s+1,\text{ }\lambda \in \mathbf{C}_{1},\text{ }|\lambda
|\rightarrow
\infty .
\end{equation*}
This and (\ref{26aa}) imply that
\begin{equation*}
e^{i\sqrt{\lambda }x}P(\lambda )+e^{-i\sqrt{\lambda }x}[S(\lambda
)-z_{0}(\lambda )]=e^{-i\sqrt{\lambda }x}O(|\lambda |^{-\infty }).
\end{equation*}
We substitute here two different values for $x$ from the interval
$(s,s+1)$
and solve the system for $P(\lambda ),$ $S(\lambda )-z_{0}(\lambda ).$
This
leads to
\begin{equation*}
S(\lambda )=z_{0}(\lambda )+O(|\lambda |^{-\infty }),\text{ \ \
}\lambda \in
\mathbf{C}_{1},\text{ }|\lambda |\rightarrow \infty .
\end{equation*}
Hence,
\begin{equation*}
\det S(\lambda )=\det z(x,\lambda )+O(|\lambda |^{-\infty })\text{ \ \
for
any }x\in (s,s+1),
\end{equation*}
and from (\ref{4s}), (\ref{5s}) it follows that
\begin{equation}
\ln [\det S(\lambda )]\sim \overset{\infty }{\underset{m=1}{\sum
}}[(2i\sqrt{ \lambda })^{-m}\int_{0}^{s}Q_{m}(x)dx]\sim
\overset{\infty }{\underset{m=1}{ \sum }}[(2i\sqrt{\lambda
})^{-m}\int_{0}^{\infty }Q_{m}(x)dx] \label{lds}
\end{equation}
as $\lambda \in \mathbf{C}_{1},$ $|\lambda |\rightarrow \infty .$
In order to complete the proof of Theorem \ref{t8} it remains only
to show that the terms with even $m$ in the formula above are
zeroes. It will be done in the process of proving the next
theorem.

The following trace type theorem (compare \cite{mnv}, formula (34))
allows
us to estimate the density of the spectral measure of the operator $H$
through the coefficients $I_{m}$ in the asymptotic expansion
(\ref{ds}).

\begin{theorem}
\label{t9}For any $m\geq 0,$
\begin{equation}
\frac{1}{\pi }\int_{0}^{\infty }\lambda ^{m-1/2}\ln |\det S(\lambda
)|d\lambda =I_{m}+(-1)^{m}\overset{q}{\underset{j=1}{\sum
}}\frac{2|\lambda
_{j}|^{m+1/2}}{2m+1}.  \label{ft9}
\end{equation}
Here $\lambda _{j}<0,$ $1\leq j\leq q,$ are eigenvalues of $H$.
\end{theorem}

\textbf{Proof}. It will be convenient for us to introduce
$z=\sqrt{\lambda } . $ Then the upper half plane $\mathrm{Im}z>0$
corresponds to $\mathbf{C} ^{\prime }=\mathbf{C}\backslash \lbrack
0,\infty ),$ and points $ z_{j}=ik_{j},$ $k_{j}=\sqrt{|\lambda
_{j}|},$ on the positive part of imaginary axis correspond to the
eigenvalues of $H$. Let $B(z)$ be the Blashke product:
\begin{equation*}
B(z)=\overset{q}{\underset{j=1}{\Pi }}\frac{z-ik_{j}}{z+ik_{j}}.
\end{equation*}
Then the function
\begin{equation*}
\ln \frac{\det S(z^{2})}{B(z)}
\end{equation*}
is analytic in the upper half plane. Since $|B(z)|=1$ for real $z,$
then
\begin{equation*}
\mathrm{Re}[\ln \frac{\det S(z^{2})}{B(z)}]=\ln |\det
S(z^{2})|,\text{ \ \ } z\in R,
\end{equation*}
and by the Herglotz formula
\begin{equation}
\frac{1}{\pi i}\int_{-\infty }^{\infty }\frac{\ln |\det
S(k^{2})|}{z-k} dk=\ln [\det S(z^{2})]-\ln B(z),\text{ \ \
}\mathrm{Im}z\geq 0. \label{111}
\end{equation}
We are going to write the asymptotic expansion for all terms in
(\ref{111})
as $\mathrm{Im}z\geq 0,$ $|z|\rightarrow \infty .$ The expansion of
the
first term in the right-hand side is given by (\ref{lds}). Obviously,
\begin{equation*}
\ln \frac{z-ik}{z+ik}\sim i\overset{\infty }{\underset{m=1}{\sum
}}\frac{ 2(-1)^{m+1}k^{2m+1}}{(2m+1)z^{2m+1}},\text{ \ \
}\mathrm{Im}z\geq 0,\text{ \ }|z|\rightarrow \infty .
\end{equation*}
This leads to the asymptotic expansion for the second term in the
right-hand side of (\ref{111}). One has to be careful when
$S(k^{2})$ is considered for real $k=k_{0}.$ The values of
$S(k_{0}^{2})$ are understood as $ S((k_{0}+i0)^{2}).$ One can
easily show using (\ref{26aa}), that $S(\lambda
+i0)=\overline{S}(\lambda -i0)$ for $\lambda >0,$ i.e.
$S(k_{0}^{2})$ is an even function. Hence, if $\mathrm{Im}z\geq
0,$ $|z|\rightarrow \infty ,$ then
\begin{eqnarray*}
\int_{-\infty }^{\infty }\frac{\ln |\det S(k^{2})|}{z-k}dk &\sim
&\overset{ \infty }{\underset{m=1}{\sum
}}\frac{1}{z^{2m+1}}\int_{-\infty }^{\infty
}k^{2m}\ln |\det S(k^{2})|dk \\
&\sim &\overset{\infty }{\underset{m=1}{\sum }}\frac{1}{z^{2m+1}}
\int_{0}^{\infty }\lambda ^{m-1/2}\ln |\det S(\lambda )|d\lambda .
\end{eqnarray*}

We equate the coefficients in the asymptotic expansions of the
left and right-hand sides of (\ref{111}) as $\mathrm{Im}z\geq 0,$
$|z|\rightarrow \infty .$ The corresponding equalities for odd
powers of $1/z$ imply (\ref {ft9}), and the equalities for even
powers of $1/z$ show that the terms in (\ref {lds}) with even $m$
are zeroes. The proofs of Theorems \ref{t8} and \ref{t9} are
completed.

The following statement is an obvious consequence of Lemma \ref{n1}
and
Theorems \ref{tes} and \ref{t9}.

\begin{theorem}
\label{t10}Let $\mu _{ac}(d\lambda )=h(\lambda )d\lambda $ be the a.c
component of the spectral measure of the operator $H$ with a matrix
potential $v\in C_{0}^{\infty }(R),$ and let $\delta =[\lambda
_{1},\lambda
_{2}],$ $\lambda _{2}>\lambda _{1}>0,$ be an interval of the positive
semiaxis. Then for any $m\geq 0$
\begin{equation*}
||[h(\lambda )]^{-1}||\geq \pi \sqrt{\lambda _{1}},\text{ \ }\lambda
\in
\delta ,
\end{equation*}
and
\begin{equation*}
\int_{\delta }\ln ||[h(\lambda )]^{-1}||d\lambda \leq \pi \lambda
_{2}^{m-1/2}\{|I_{m}|+\overset{q}{\underset{j=1}{\sum
}}\frac{2|\lambda
_{j}|^{m+1/2}}{2m+1}\}.
\end{equation*}
\end{theorem}

The last statement, which will be proven in this section, is devoted to
the
generalization of the Lieb-Thirring estimates to the matrix case. It
will
allow us to simplify the estimate in Theorem \ref{t10}.

\begin{lemma}
\label{llt}Let $\{\lambda _{j}\}$ be the set of negative eigenvalues
for the
Schr\"{o}dinger operator with a matrix potential $v,$ and let
\begin{equation*}
v_{\gamma }:=\int_{0}^{\infty }||v(x)||^{\gamma +1/2}dx<\infty .
\end{equation*}
Then
\begin{equation*}
\underset{j}{\sum }|\lambda _{j}|^{\gamma }\leq nv_{\gamma },
\end{equation*}
where $n\times n$ is the size of the matrix $v$.
\end{lemma}

\textbf{Proof}. Let $\mu (x)=||v(x)||$ and let
$H_{-}=-\frac{d^{2}}{dx^{2}}-$ $\mu (x)I$ be the matrix
Schr\"{o}dinger operator on $L_{2}(R_{+})$ with the same b.c. at
$x=0$ as for operator $H$. Then $H_{-}\leq H$ and $\lambda
_{i}^{-}\leq \lambda _{i},$ where $\lambda _{i}^{-}$ are
eigenvalues of $H_{-}$ and the eigenvalues $\{\lambda _{i}^{-}\}$
and $\{\lambda _{i}\}$ are numerated in increasing order. Thus,
\begin{equation*}
\underset{j}{\sum }|\lambda _{j}|^{\gamma }\leq \underset{j}{\sum
}|\lambda
_{j}^{-}|^{\gamma }.
\end{equation*}
It remains only to note that $\{\lambda _{i}^{-}\}$ are eigenvalues of
the
scalar Schr\"{o}dinger operator repeated $n$ times.

Let us recall that the functional $J_{m}$ (see (\ref{jp})) has the same
form
as the functional $I_{m}$ (see (\ref{pm})) with the integrand in
(\ref{jp})
being a generalized homogeneous of order $2m+2$ function of $v$ and
its
derivatives. There exists an independent of $v$ constant $c_{m}$ such
that
\begin{equation*}
|I_{m}|\leq c_{m}J_{m}.
\end{equation*}
This estimate was proven in the scalar case in \cite{mnv} using
Kolmogorov type estimates, and the same proof remains valid in the
matrix case. The last estimate together with Lemma \ref{llt} allow
us to rewrite Theorem \ref{t10} in the following form.

\begin{theorem}
\label{t11}Let $\mu _{ac}(d\lambda )=h(\lambda )d\lambda $ be the
a.c component of the spectral measure of the operator $H$ with a
matrix potential $v\in C_{0}^{\infty }(R),$ and let $\delta
=[\lambda _{1},\lambda _{2}],$ $\lambda _{2}>\lambda _{1}>0,$ be
an interval of the positive semiaxis. Then there exist independent
of $v$ constants $c(\delta )>0$ and $ C(\delta )$ such that for
any $m\geq 0$
\begin{equation*}
||[h(\lambda )]^{-1}||\geq c(\delta ),\text{ \ }\lambda \in \delta
,\text{\
\ and \ \ }\int_{\delta }\ln ||[h(\lambda )]^{-1}||d\lambda \leq
C(\delta
)J_{m}(v).
\end{equation*}
\end{theorem}

\textbf{5. Generalized }$L_{2}$\textbf{-conjecture.} We shall need the
following

\begin{lemma}
\label{ll}Let $v=v(x)$ be a matrix potential such that
\begin{equation*}
J_{p}(v)<\infty
\end{equation*}
for some $p\geq 0,$ where $J_{p}$ is the functional defined in
(\ref{jp}).
Then there is a sequence of matrix potentials $v_{s}\in C_{0}^{\infty
}(R_{+})$ such that $v_{s}(x)=0$ for $x>s$ and
\begin{equation}
J_{p}(v-v_{s})\rightarrow 0,\text{ \ \
}\int_{0}^{s}||v(x)-v_{s}(x)||dx \rightarrow 0\text{ \ \ as \
}s\rightarrow \infty .  \label{re}
\end{equation}
\end{lemma}

\textbf{Proof}. We fix a function $\varphi _{s}(x)\in C^{\infty }(R),$
such
that $\varphi _{s}(x)=1$ for $x<s-1,$ $\varphi _{s}(x)=0$ for
$x>s-1/2.$ The
boundedness of $J_{p}$ implies that
\begin{equation*}
\int_{s-1}^{\infty
}(||v^{(p-1)}(x)||^{2}+||v(x)||^{p+1})dx\rightarrow 0 \text{ \ \
as \ }s\rightarrow \infty .
\end{equation*}
From here it follows that relations (\ref{re}) hold for the matrix
$ v-\varphi _{s}v.$ Let $\alpha =\alpha (x)\ $be a$\ C_{0}^{\infty
}(R)$ -function with the support strictly inside of the interval
$[0,1].$ Let
\begin{equation*}
\alpha _{\varepsilon }(x)=\varepsilon ^{-1}\alpha (\varepsilon
x)/\int_{0}^{1}\alpha (x)dx.
\end{equation*}
The convolution $v_{s,\varepsilon }=\varphi _{s}v\ast \alpha
_{\varepsilon }$ is a$\ C_{0}^{\infty }(R)$-function supported on
the interval $(0,s)$ if $ \varepsilon <1/2.$ Since
\begin{equation}
J_{p}(\varphi _{s}v-v_{s,\varepsilon })\rightarrow 0,\text{ \ \ }
\int_{0}^{s}||\varphi _{s}v(x)-v_{s,\varepsilon
}(x)||dx\rightarrow 0\text{ \ \ as \ }\varepsilon \rightarrow 0,
\label{rea}
\end{equation}
then one can choose $\varepsilon =\varepsilon (s)<1/2$ to be so
small that the left-hand sides of (\ref{rea}) are less than $1/s.$
Then $ v_{s}=v_{s,\varepsilon (s)}$ satisfies the requirements of
Lemma \ref{ll}. The proof is completed.

The following criteria of the absolute continuity of the limit measure
was
proven in \cite{mnv}.

\begin{theorem}
\label{tmnv}Let $\mu _{s}(d\lambda )=h_{s}(\lambda )d\lambda $ be
a sequence of absolutely continuous positive (scalar) measures on
an interval $\delta =[\lambda _{1},\lambda _{2}].$ Let $\mu
_{s}(d\lambda )\ $\ converge weakly to a measure $\mu (d\lambda )$
as $s\rightarrow \infty ,$ and for any $s$
\begin{equation}
\underset{\delta \cap \{h(\lambda )<1\}}{\int }\ln
\frac{1}{h_{s}(\lambda )} d\lambda <C<\infty ,  \label{mnv}
\end{equation}
where $C$ does not depend on $s$. Then the essential support of the
a.c.
component $\mu _{ac}(d\lambda )$ of the limit measure coincides with
$\delta
.$
\end{theorem}

\textbf{Remarks 1}. If $h_{s}(\lambda )<c<\infty $ then one can
integrate
over $\delta $ in (\ref{mnv}), since the integral over $\delta \cap
\{h(\lambda )>1\}$ is uniformly bounded in this case.

\textbf{2}. It was also shown in \cite{mnv} that the statement of the
theorem remains valid if the logarithmic function above is replaced by
any
function monotonically increasing on $[0,\infty ]$ (we shall not need
this
fact).

We shall say that an a.c. $n\times n$ matrix measure $\mu (d\lambda
)=\nu
(\lambda )d\lambda $ has the multiplicity \ $n$ and is essentially
supported
on some interval $\delta =[\lambda _{1},\lambda _{2}]$ if the latter is
true
for all scalar measures $\nu _{j}(\lambda )d\lambda $ where $\nu
_{j}(\lambda )$ are eigenvalues of the matrix $\nu (\lambda ).$ It is
easy
to see that the following matrix analogue of Theorem \ref{tmnv} can be
immediately reduced to a scalar case.

\begin{theorem}
\label{last}Let $\mu _{s}(d\lambda )=h_{s}(\lambda )d\lambda $ be
a sequence of absolutely continuous positive matrix measures on an
interval $\delta =[\lambda _{1},\lambda _{2}].$ Let $\mu
_{s}(d\lambda )\ $ converge weakly to a measure $\mu (d\lambda )$
as $s\rightarrow \infty ,$ and for any $s$
\begin{equation}
||[h_{s}(\lambda )]^{-1}||\geq c(\delta )>0,\text{ \ \ }\lambda
\in \delta ; \text{ \ \ \ }\int_{\delta }\ln ||[h_{s}(\lambda
)]^{-1}||d\lambda <C(\delta )<\infty ,  \label{mnv1}
\end{equation}
where $c,C$ do not depend on $s$. Then the a.c. component $\mu
_{ac}(d\lambda )$ of the limit measure has multiplicity $n$ and is
essentially supported on $\delta .$
\end{theorem}

The following theorem presents the main result.

\begin{theorem}
\label{tl}Let $H$ be a Schr\"{o}dinger operator with a matrix
potential $v=v(x)$ such that
\begin{equation*}
J_{p}(v)<\infty
\end{equation*}
for some $p\geq 0.$ Then the spectral measure $\mu (d\lambda )$ of the
operator $H$ has the a.c. component of the multiplicity \ $n$ which is
essentially supported on $[0,\infty ).$
\end{theorem}

\textbf{Proof.} Let $\{v_{s}\}$ be a sequence of $C_{0}^{\infty
}(R_{+})$ -potentials constructed in Lemma \ref{ll} and let $\mu
_{s}(d\lambda )$ be the spectral measure of the operator $H_{s}$
with the potential $v_{s}.$ Lemma \ref{nlsm} and the second of
relations (\ref{re}) imply the weak convergence of $\mu
_{s}(d\lambda )$ to $\mu (d\lambda )$ as $s\rightarrow \infty .$
Hence, the convergence holds for restrictions of these measures on
the semiaxis $\lambda >0,$ where measures $\mu _{s}(d\lambda )$
are a.c. (see Lemma \ref{n1}). Due to the first of relations
(\ref{re}), $ J_{p}(v_{s})\leq C=2J_{p}(v)$ if $s$ is big enough.
Hence, Theorem \ref{tl} is an immediate consequence of Theorems
\ref{t11} and \ref{last}.

\end{document}